\documentclass[12pt]{article}
\usepackage{graphicx}
\usepackage{fullpage}

\def\be{\begin{equation}}
\def\ee{\end{equation}}
\def\bea{\begin{eqnarray}}
\def\eea{\end{eqnarray}}

\usepackage[scaled=0.92]{helvet}	

\usepackage{amsmath}		

\begin{document}

\title{Piecewise Linear Phase Transitions}         
\author{Joseph B.\ Keller\\Departments of Mathematics and Mechanical Engineering\\Stanford University}        
\date{\today}          
\maketitle

\begin{abstract}
It is shown how simple assumptions lead to piecewise linear behavior, which is observed in certain phase transitions.
\end{abstract}


In certain phase transitions, some property $F(\varphi)$ of a system is a piecewise linear function of a parameter $\varphi$.  Its slope $s(\varphi)=dF(\varphi)/d\varphi$ changes discontinuously at a critical value $\varphi_c$ of $\varphi$.   An example is provided by the force $F(\varphi)$ required to push a rod into a layer of glass beads of volume fraction $\varphi$, $.57 <\varphi < .63$, at constant velocity (Schr\"oter et al., \cite{schroter}).  When the penetration depth is 60mm, $F(\varphi)$ is nearly piecewise linear, with its slope changing nearly discontinuously at $\varphi_c = .60$.  ( See Figure 1.)  We shall show that simple assumptions lead to piecewise linear behavior of $F(\varphi)$.  Similar assumptions lead to the sharp transition from one power law to another power law, which occurs in turbulent boundary layer flow.  \cite{keller1}.

Suppose the derivative of the slope $s(\varphi)$ is a smooth function $f(s, \varphi)$, which we expand in a Taylor series about some value $s_0$, through quadratic terms. Then
\be
\frac{ds(\varphi)}{d\varphi} = f(s, \varphi) =f(s_0, \varphi) + f_s (s_0, \varphi) (s-s_0) +\frac{1}{2} f_{ss} (s_0 , \varphi) (s-s_0)^2 +O \left[ \left( s-s_0\right)^3\right].
\label{1}
\ee
We write the quadratic polynomial in $s$, on the right side of (\ref{1}),  in terms of its roots $\alpha$ and $\beta$. We assume them to be real, with $\alpha<\beta$:
\be
\frac{ds(\varphi)}{d\varphi} =\gamma (s-\alpha) (\beta-s) +O \left[ \left( s-s_0\right)^3\right].
\label{2}
\ee

When we omit the error  term in (\ref{2}), and approximate $\alpha, \beta$ and $\gamma$ by constants, we can solve the resulting differential equation for $s(\varphi)$.   Upon setting $s(\varphi_c)= (\alpha +\beta)/2$, we get
\be
s(\varphi) =
\frac{\alpha + \beta e^{(\beta-\alpha) \gamma (\varphi -\varphi_c)}}
{1+e^{(\beta-\alpha )\gamma (\varphi -\varphi_c)}}.
\label{3}
\ee
As $\gamma\to \infty$, the slope $s(\varphi)$ given by (\ref{3}) becomes discontinuous at $\varphi_c$, tending to $\alpha$ for $\varphi < \varphi_c$ and to $\beta$ for $\varphi>\varphi_c$.  For $\gamma$ finite but large, $s(\varphi)$ changes continously but rapidly from $\alpha$ to $\beta$.

$F(\varphi)$ is obtained by integrating the defining equation $dF/d \varphi=s(\varphi)$, with $s$ given by (\ref{3}):
\bea
F(\varphi) &=& F(\varphi_c) +\int^\varphi_{\varphi_c} s(\varphi^\prime) d\varphi^\prime\nonumber\\
&=& F(\varphi_c) +\beta (\varphi-\varphi_c) +\frac{1}{\gamma} \left[ \log \left( 1+e^{(\beta-\alpha) \gamma (\varphi-\varphi_c)} \right) -\log 2\right].
\label{4}
\eea
As $\gamma \to \infty$, $F(\varphi)$ given by (\ref{4}) tends to the piecewise linear function
\bea
\lim\limits_{\gamma\to \infty} F(\varphi)&=& F(\varphi_c) + \alpha (\varphi -\varphi_c), \qquad \varphi< \varphi_c,\nonumber\\
&=& F(\varphi_c) +\beta (\varphi-\varphi_c) , \qquad \varphi> \varphi_c.
\label{5}
\eea
This is the form of $F(\varphi)$ shown in Figure 1 and in Figure 2 of \cite{schroter}.  For $\gamma$ finite, (\ref{4}) provides a smooth approximation to the piecewise linear function in (\ref{5}).

In \cite{keller1} we treated the mean velocity $u(y)$ at distance $y$ from a wall in a  turbulent boundary layer flow.  We defined the dimensionless shear $s(y) =d\log u/d\log y$, and proceeded as in (\ref{1}) -- (\ref{3}) with $\varphi =y$ and $\varphi_c =y_c$.  Then we integrated the defining equation $d\log u/d\log y= s(y)$, with $s(y)$ given by (\ref{3}), to obtain $u(y)$.  As $\gamma\to \infty$, the result (\ref{3}) for $s(y)$ tends to $\alpha$ for $y<y_c$ and to $\beta$ for $y>y_c$.  The result for $u(y)$ tends to $Ay^\alpha$ for $y<y_c$ and to $By^\beta$ for $y>y_c$, with $A$ and $B$ constants.  Previously these two power laws had been  shown to fit the experimental data by Barenblatt, Chorin, and Prostokoshin \cite{bcp}.

\newpage

\vskip.25in
\begin{figure}[htbp]\begin{center}
\includegraphics[width=0.75\textwidth]{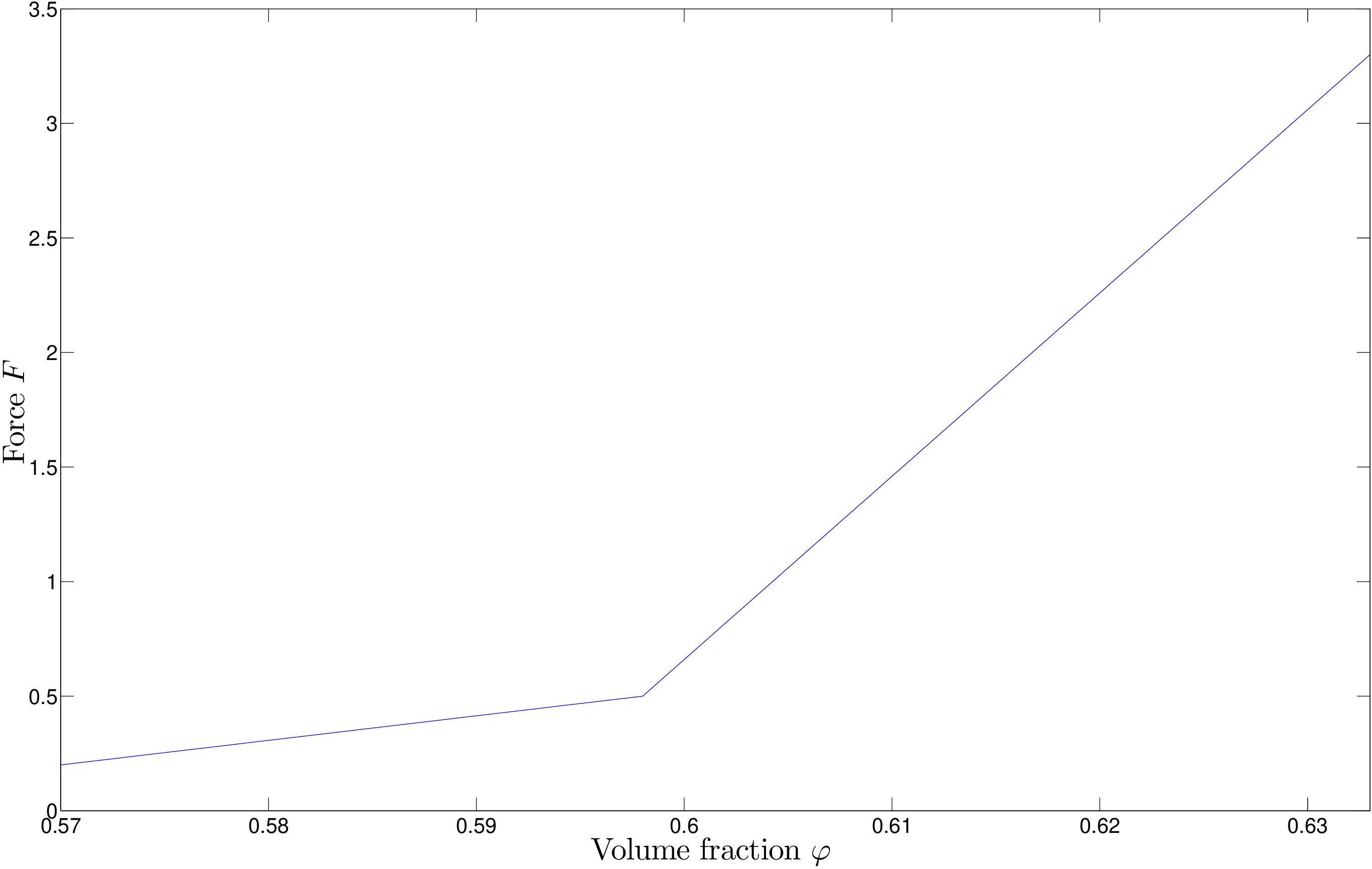}
\end{center}\end{figure}

\vskip.25truein


Figure 1.  $F(\varphi)$ versus $\varphi$ given by (\ref{5}) with $\varphi_c = .598$, $F(\varphi_c)=.5$, $\alpha =10.7$ and $\beta =80$.  These values were chosen to make the graph resemble that in Figure 2 of \cite{schroter}, which shows the force $F(\varphi)$ in newtons versus the volume fraction $\varphi$ for a penetration depth of 60mm.


\begin{thebibliography}{99}

\bibitem{schroter}M.\ Schr\"oter, S.\ Nagle, C.\ Radin, and H.\ L.\ Swinney, Phase transition in a static granular system, Europhysics Letters 78 (2007) 44004; arXiv: cond-mat/0606459v2 [cond-mat.soft].

\bibitem{keller1}J.\ B.\ Keller, ``Power laws for turbulent boundary layer flow,'' {\em Phys.\ Fluids} {\bf 14}, L89 (2002).

\bibitem{bcp}G.\ I.\ Barenblatt, A.\ J.\ Chorin, and V.\ M.\ Prostokoshin, ``A note on the intermediate region in turbulent boundary layers,'' {\em Phys.\ Fluids} {\bf 12}, 2159 (2000).

\end{thebibliography}
\end{document}